# A Novel Approach in Calculating Stakeholder priority in Requirements Elicitation

Anupama Prasanth
College of Computer Studies
[3]AMA International University
Kingdom of Bahrain
anupama.prasanth@amaiu.edu.bh

Sandhia Valsala
College of Business
[2]University of Buraimi
Sultanate of Oman
sandhia.v@uob.edu.om

Safeeullah Soomro
College of Computer Studies
[3]AMA International University
Kingdom of Bahrain
s.soomro@amaiu.edu.bh

*Abstract*—The ultimate goal of any software developer seeking a competitive edge is to meet stakeholders' needs and expectations. To achieve this, it is necessary to effectively and accurately manage stakeholders' system requirements. The paper proposes a systematic way of classifying stakeholders and then describes a novel method for calculating stakeholder priority taking into consideration the fact that different stakeholders will have different importance level and different requirement preference. Finally the requirement preference calculation is done where stakeholders choose the best requirements based on two factors, value and urgency of the requirement. The proposed method actively involves stakeholders in the requirement elicitation process (*Abstract*)

## I. INTRODUCTION (*Heading 1*)

One of the most important activities during requirements engineering is requirement elicitation i.e.to select and prioritize those requirements that satisfies various explicit and implicit objectives and constraints and to remove the unimportant ones. [9]. Removing unwanted requirement and accepting the most suitable and appropriate requirement is one of the most important step in requirements engineering [6]. There should be a proper balance between system needs and user needs. An ineffective elicitation process results in the inclusion of unwanted requirements which could delay the project and also could over budget the project. The requirements that need to be included in the project should satisfy various constraints like stakeholder preference, resources, cost of development etc. It is impossible to have all the available requirements implemented, hence there is a need to prioritize and choose only those requirements satisfying various technical and non-technical constraints [10].

Involving stakeholders in requirement engineering process is considered to be the biggest challenge of software development. System stakeholders in the area of software engineering are defined as "people or organizations who will be affected by the system and who have a direct or indirect influence on the system requirements. [17]. A successful system needs to satisfy the interest of different group of stakeholders

Success of a system depends on identifying the correct requirements and in order to know the requirements we need to identify the stakeholder's needs and desires. If the correct group of stakeholders are not included then it may lead to the inclusion of irrelevant requirements which could ultimately lead to system failure [13].On the other hand if stakeholders are not involved, requirements become incomplete and at the end projects will fall under challenged category. Missing stakeholders implies missing essential requirements which subsequently increase project costs and causes project delay [3].

One of the most important step in requirement engineering activity is Stakeholder identification i.e. determining who the stakeholders are and how important they are. The paper proposes a novel method for calculating stakeholder priority as all the identified stakeholders will not have the same priority or importance. Also the stakeholders will have different preference in implementing the requirements of a system. The proposed method allows stakeholders to choose the requirements to be implemented in a system based on two factors namely urgency and value of the requirement. The proposed method is tested on a sample data set and the results are recorded.

The organization of this paper is as follows. Section 2 analyzes the reasons that can contribute to failure of a project; Section 3 explains the existing methods of classifying stakeholders and stakeholder priority calculation adopted by various existing release planning models .Section 4 and 5 explains our proposed method of calculating stakeholder priority and requirement preference calculation followed by conclusion and future work

## II. REASONS OF SOFTWARE FAILURE

In Chaos Report (Standish Reports) there are several reasons that can contribute to the failure of a project, some of the major factors being incomplete requirements, lack of stakeholder involvement and delay in project delivery. On analyzing the Chaos Standish Repot (2010-2015) as given in the Table 1 given below shows that projects falling in the challenged category shows a steady increaseMaintaining the Integrity of the Specifications.

Table 1: Standish Chaos Report (Standiah Chaos Report, 2010-2015)

| Year | Project Succeeded | Project Failed | Project Challenged |
|------|-------------------|----------------|--------------------|
| 2010 | 37% | 21% | 42% |
| 2011 | 29% | 22% | 49% |
| 2012 | 27% | 17% | 56% |
| 2013 | 31% | 19% | 50% |
| 2014 | 28% | 17% | 55% |
| 2015 | 29% | 19% | 52% |

The main factor that contributes to project failure or projects falling into challenged category was the lack of stakeholder/user involvement. From the data published by the Standish group (Chaos Report 2014) it is clear that 12.3% of the projects are challenged due to incomplete requirements and specifications ,12.8 % due to lack of user involvement and 4.3% due to unrealistic time frames. From these findings it can also be concluded that 30% of the projects fail due to lack of stakeholder involvement. Not only involving the stakeholders is important for project success but it is equally important to select the most appropriate stakeholders [15]. To meet the goals of a software project its necessary to understand the stakeholders and assure their involvement in identifying the most appropriate requirements/features to be included in a system.

For example, the RE process may involve people who are connected with the system in one way or another which could be clients, developers, users and customers. On the other hand, it can also include the development team who elicit, design and construct the system as well as the system users who use the system to fulfill their daily tasks. Different stakeholders will have different roles, influence and positions; it's a very important factor to be considered during requirement elicitation process. One should also not forget the fact that the projects are bound to tight schedule and limited budget [16].

From the Standish report analysis it is clear that one of the major factor contributing to project failure is the lack of stakeholder/user involvement and the inclusion of inappropriate requirements/features in the system. Hence it is important to understand the role of stakeholders in RE process, equally important is classifying the stakeholders into appropriate groups as all the stakeholders involved in system development process will not have the same priority or importance.

It is important to classify these stakeholders as effective participation of stakeholders in a project subsequently improves requirements quality and thus reduces the chances of project failure.

## III. STAKEHOLDER CLASSIFICATION AND STAKEHOLDER PRIORITY – RELATED STUDIES

The ultimate goal of any software developer seeking a competitive edge is to meet stakeholders' needs and expectations. To achieve this, it is necessary to effectively and accurately manage stakeholders' system requirements. Software development has become a highly complicated and critical activity in today's world Gone are those days where people use whatever systems are developed for them but now people reject those that do not cater to their needs .The user wants to see their wish list completely implemented in the systems they use. The challenges of software development can be reduced a lot if the correct groups of stakeholders are participating in the requirement elicitation process. It is a fact that different stakeholders prefer different requirements over others. System stakeholders in the area of software engineering are defined as "people or organizations who will be affected by the system and who have a direct or indirect influence on the system requirements" [17]. Now the main question that arises here is how to classify the stakeholders? Project success depends on the inclusion of the most appropriate, correct requirements and this can be achieved only if we have the right group of stakeholders. Therefore, it is an essential task to identify and choose the most appropriate stakeholders.

Stakeholder categorization is the core of stakeholder identification. Identifying the most appropriate stakeholders is an important task. The developed [12] theory of stakeholder identification is done through an analysis of „what" and „who" affects the organization. They classifies stakeholder into three categories namely (a) latent or low salience, (b) expectant or moderately salient, and (c) definitive or highly salient. An approach [20] to discover all stakeholders in a specific software project development by establishing a set of stakeholder denominated "baselines" as supplier, client and satellite stakeholder. Another developed [7] method was to identify stakeholders and their different viewpoints within a computer information system using a legacy system. The purpose of MEWSIC [23] was to provide software developers with a practical tool to identify stakeholders. The method grouped all the people involved in a project depending on the priorities of their interests. McMenu [11] classifies stakeholders as primary, secondary, external, and extended. Another classification as given by Preis classifies stakeholders into goal oriented and mean oriented. [14].

Most of these requirements or features that need to be considered in software arise from diverse stakeholders but to include all these requirements is also practically impossible. So deciding on which feature (requirement) should go to the software is a crucial decision considering the fact that only a subset of requirements could only be implemented and the selection of the most appropriate and crucial requirements lies in the hand of stakeholders. One of the most important factors in deciding which requirements have to be included is based on Stakeholder's decision. Effectively solving the RP problem involves satisfying the needs of these stakeholders. Examples of stakeholder' are: users, managers, developers, sales representatives etc. Most of the existing releases planning models consider stakeholder priority or preference in selecting the requirements. On reviewing the existing software release planning models [22] it can be clearly seen that 16 out of the identified 31 models consider stakeholder priority as one of the deciding factor for selecting requirements during RP process.

Most of the existing RE models considers stakeholder's decision as a factor in deciding the requirement to be included during the software development process. However only few models concentrate on calculating stakeholder priority taking into consideration the fact that all the stakeholders will not have the same importance. The method [24] includes those requirements what the stakeholder considers the most valuable. This model [8] considers stakeholders decision in selecting requirements and assigns a stakeholder weight ranging from 1-5 based on their importance in the company. The EVOLVE series [17], [19]attempts to consider stakeholder preference in determine the most appropriate requirements but just assigns a numerical weight to each stakeholder depending on their position in the company. On analyzing the 32 models involved in RP process [22] only 55% of the models consider stakeholders' influence in requirements selection. No attempt is made by any of these models to group the stakeholders based on their importance. Most of these models [1], [2], [4], [5] attempt to assign a numeric weight to the stakeholders based on the position in their company. The proposed solution overcomes this drawback by properly grouping them and then assigning a weight to each one of them as within a group also the stakeholders may have different weightings.

## IV. STAKEHOLDER PRIORITY CALCULATION – A NOVEL APPROACH

A diverse group of stakeholders will be involved and it is essential to satisfy the needs of these stakeholders in order to solve the RP problem effectively. Adopting the model of stakeholder classification proposed by [12] the stakeholders are classified into 3 categories namely (a) latent or low salience, (b) expectant or moderately salient, and (c) definitive or highly salient.

Each stakeholder group, potential or actual, has a certain level of importance for the company, depending on the attributes it possesses. Mitchell, Agle and Wood came up with three main categories of stakeholders' attributes: (1) *power*– stakeholders possess power to influence the company; (2) *legitimacy*– stakeholders have legitimate relationship with the company; and (3) *urgency*– stakeholders have urgent claim on the company [12]. Based on these attributes, the authors identified different types of stakeholders, possessing one or more attributes. Stakeholder salience model has stakeholder classes which are separated in three main groups- *latent stakeholders*– those stakeholder groups who possess only one of the three attributes of power, legitimacy and urgency; *expectant stakeholders*– those groups who possess two attributes; and *definitive stakeholders*– those who possess all three attributes [12].

We adopt the Stakeholder Salience Theory proposed by Mitchell, Agle and Wood in classifying the stakeholders. Stakeholders are initially verified to check the attribute they process. Some stakeholders will possess only one attribute some may satisfy two attribute and some others may satisfy all the three attributes. Those stakeholders that possess only one attribute are placed under latent or low salience group and those who possess two attributes are placed under expectant or moderately salient group and those who possess all the three attributes are placed under definitive or highly salient group [12].

Let's give a weight to these three groups 3 for highly salient group , 2 for moderately salient group and 1 for low salient group. In each group there will multiple stakeholders involved. All the stakeholders within the group will not have the same importance, so it is necessary to calculate the weight of each stakeholder under each group and the average of the stakeholder weight will give the stakeholder priority of that group. Then we multiply the stakeholder priority of that group with the weight of the group to get the final stakeholder priority of that group.

The sample dataset consists of 10 stakeholders and 8 requirements. The stakeholders are identified as s1,s2…s10.As per the initial grouping s1,s4,s5 belongs to low salient group , s2,s3,s7,s9 belongs to moderately salient group and s6,s8,s10 belongs to highly salient group as depicted in the **Table 2** given below

**Table 2: Stakeholder-Groups**

| Low Salient group | Moderate Salient Group | Highly Salient Group |
|---|---|---|
| s1,s4,s5 | s2,s3,s7,s9 | s6,s8,s10 |

Within the group all the stakeholders will not have the same priority. So the stakeholders in each group are weighted using AHP from a general project management perspective .A Matrix of pair-wise comparison of stakeholders on a nine-point scale is done based on the given below. Weightings are used to discriminate between stakeholders. For greater flexibility and objectivity these weightings are further calculated using the pairwise comparison method from AHP. [18].

**Table 3: Fundamental Scale of absolute Numbers** (Saaty & Sodenkamp, 2010)

| Numerical Values | Verbal judgement |
|---|---|
| 1 | Equal Importance |
| 2 | Weak or Sight |
| 3 | Moderate Importance |
| 4 | Moderate Plus |
| 5 | Strong Importance |

| 6 | Strong Plus |
| 7 | Very Strong |
| 8 | Very , Very Strong |
| 9 | Extreme Importance |
| Reciprocals of above : If activity I has one of the above non-zero numbers assigned to it when compared with activity j, then j has the reciprocal value when compared with i | |

After representing the comparison matrix, priorities are computed by finding the principal Eigen values and the corresponding Eigen vector of the pairwise comparison matrix for each stakeholder group. The normalized principal Eigen vector is the priorities vector. To compute the priorities the following two steps are to be carried out [8].

The stakeholder weightings are computed from the eigenvalues of the matrix shown below. The computed stakeholder weightings is shown in the table. which shows the representation for low salient group. The technique of averaging over normalized columns can be used to approximate the eigenvalues. The normalized principal Eigen vector is also called priority vector. Since it is normalized, the sum of all elements in priority vector is 1. The priority vector shows relative weights among the considered stakeholders. After calculating individual stakeholder weight we average it to find the stakeholder priority of that group. The final stakeholder weightage of each group is calculated by multiplying the group weight with the calculate stakeholder priority of that group.

**Table 4: Matrix of pair-wise comparison of stakeholders on a nine point scale**

| Low-Salient Group | s1 | s4 | s5 |
|---|---|---|---|
| s1 | 1 | 2 | 3 |
| s4 | 1/2 | 1 | 4 |
| s5 | 1/3 | 1/4 | 1 |

The **Table 5** given below shows the computed Stakeholder weight for the low salient group.

**Table 5 : Calculated Stakeholder weight**

| Stakeholder | Stakeholder weight |
|---|---|
| s1 | .52 |
| s4 | .36 |
| s5 | .13 |
| **AVERAGE** | **.33** |

To compute the final stakeholder priority of this group we multiply the calculated average of stakeholder weight with the group weight.

Stakeholder Priority of Group 1= Calculated Stakeholder weight * Group 1 weight

After calculating the stakeholder priority for each group, the next section summarizes the process of calculating the stakeholder requirement preference score.

V. STAKEHOLDER REQUIREMENT PREFERENCE CALCULATION

The sample dataset consists of 8 requirements (Req1,Req2…Req8) and each group of stakeholders will have different preference in selecting and implementing these requirement. Stake holder preference on implementing the requirement will be based on the two factors mainly Value (how valuable is this feature?) and Urgency (how urgent is the feature?).

Each stakeholder group will assign a value in the scale range 1-5 according to the perceived value ( ie the expected relative impact of the requirement on business value of the final

| Stake Holder Group | Req1 | Req 2 | Req 3 | Req 4 | Req 5 | Req 6 | Req 7 | Req8 |
|---|---|---|---|---|---|---|---|---|
| Group-1 (Low Salient Group) | 4 | 3 | 2 | 5 | 3 | 2 | 1 | 4 |
| Group-2(ModerateSalient Group) | 5 | 3 | 5 | 4 | 2 | 4 | 2 | 2 |
| Group-3 (High Salient Group) | 4 | 3 | 5 | 2 | 1 | 3 | 2 | 1 |

product). The scale used will be 1– no value, 2 – little value, 3 – some value, 4 – high value, and 5 – very high value" .The Table 6 given below shows the sample input of the stakeholders belonging to the three groups as identified above on their preference of requirements based on value.

**Table 6 : Stakeholder Preference based on Value of Requirement**

Final weight of each requirement is calculated by multiplying the requirement weight with the calculated stakeholder priority for each group and then summing it up. We had calculated the stakeholder priority of group-1 to be .33 similarly the stakeholder priority of group-2 to be .57and stakeholder priority of group-3 to be .75.

Weight of Requirement 1 will be 4 * .33 + 5*.57 + 4*.75 which equals to 7.17.Similarly the weight of all the other requirements are calculated.

Now the second factor to be considered is the urgency (how fast the requirement need to be implemented) of each requirement to each stakeholder. The Table 7**Table 6** given below shows the sample input of the stakeholders belonging to the three groups as identified above on their preference of requirements based on urgency.

TABLE 7: STAKEHOLDER PREFERENCE BASED ON URGENCY OF THE REQUIREMENT

| Stake Holder Group | Req1 | Req2 | Req3 | Req4 | Req5 | Req6 | Req7 | Req8 |
|---|---|---|---|---|---|---|---|---|
| Group-1 (Low Salient Group) | 3 | 4 | 5 | 1 | 2 | 3 | 4 | 2 |
| Group-2 (Moderate Salient Group) | 5 | 4 | 2 | 1 | 3 | 4 | 2 | 2 |
| Group-3 (High Salient Group) | 4 | 3 | 2 | 5 | 3 | 2 | 2 | 3 |

Final weight of each requirement is calculated by multiplying the requirement weight with the calculated stakeholder priority for each group and then summing it up.

Weight of Requirement 1 will be 3 * .33 + 5*.57 + 4*.75 which equals to **6.84**. Similarly the weight of all the other requirements are calculated.

To obtain the final requirement priority we add up the obtained calculated requirement weight based on value and the calculated requirement weight based on urgency.

**Requirement Weight = Calculated Requirement Priority based on Urgency + Calculated Requirement Priority based on Value.**

Finally we sort the requirement list based on the calculated requirement weight and the highly preferred requirements are implemented.

## CONCLUSION AND FUTURE WORKS

Stakeholders are an integral part of requirement engineering process. Not only involving the stakeholders is important but also involving the correct group of stakeholders and choosing the most appropriate requirements is also important. We have seen that one of the major reason of software failure as identified by Chaos report (Standiah Chaos Report, 2010-2015) is lack of Stakeholder involvement. So this paper proposes a method of calculating stakeholder priority by adopting the Stakeholder Salience Theory proposed by Mitchell, Agle and Wood [12] in classifying the stakeholders. After calculating the stakeholder priority; the requirement preference calculation is done where stakeholders choose the best requirements based on two factors, the value and the urgency of the requirement. The proposed method actively involves stakeholders in requirement elicitation process. The proposed methods are tested on a small dataset and results are recorded.

The research can be further extended to include additional requirement selection factors as the proposed method considers only urgency and value. This research provides solution to RP problem based on fixed estimations of parameters. We assume that these parameters are constant and do not change dynamically with environment.